\documentclass[12pt]{revtex4b3}
\usepackage{graphicx}
\usepackage{multicol}
\usepackage[ansinew]{inputenc}
\begin{document}
\bibliographystyle{apsrev}

\title{Negative phase time for Scattering at Quantum Wells:\\
       A Microwave Analogy Experiment}

\author{R.-M. Vetter, A. Haibel, G. Nimtz}

\affiliation{Universität zu Köln, II. Phys.~Institut,
             Zülpicher Str.\,77, D-50937 Köln, Germany}

\date{28 Jun 2000}

\begin{abstract}
If a quantum mechanical particle is scattered by a potential well, the wave function of the particle 
can propagate with negative phase time.
Due to the analogy of the Schr\"odinger and the Helmholtz equation this phenomenon is expected to be 
observable for electromagnetic wave propagation. Experimental data of electromagnetic wells realized 
by wave guides filled with different dielectrics confirm this conjecture now.
\end{abstract}

\maketitle

The propagation of a wave packet is determined by the dispersion relation of the medium. 
E.g. in vacuum a plane wave propagates with a constant amplitude and a phase shift 
proportional to frequency. In the case of tunnelling through a barrier, the constant 
phase leads to propagation speeds faster than light, calculated by~\cite{tunnel_theo} and measured for 
microwaves, single photons and infrared light~\cite{tunnel_exp1,tunnel_exp2,tunnel_exp3}.
In the contrary case of particles scattered by a potential well instead of a barrier, 
Li and Wang predicted a non--evanescent propagation but also with negative phase shifts~\cite{wang}. 
We present here an experimental simulation of the quantum well by a microwave set--up 
performing the analogy between the Schrödinger and the Helmholtz equation.

Applying the stationary phase approximation, the peak value of a quantum mechanical wave 
packet with a mean impulse $p_0=\hbar k_0$ propagates with the group velocity~$v_{\rm gr}=d\omega/dk|_{k_0}$. 
This relationship can also be described in terms of classical mechanics 
by $v_{\rm gr}  =  \frac{x}{\frac{d}{d\omega} kx}  =  \frac{x}{\tau}$,
where particles traverse the distance $x$ in the time $\tau$. The term $kx=\varphi$ describes the 
change of phase for the considered distance, $\frac{d}{d\omega} \, \varphi = \tau_\varphi$ is the time 
necessary for the propagation, called {\em phase time}~\cite{tunnel_theo}.

\section*{Potential scattering at quantum wells}

\begin{figure}[hbt]
  \includegraphics[width=\textwidth]{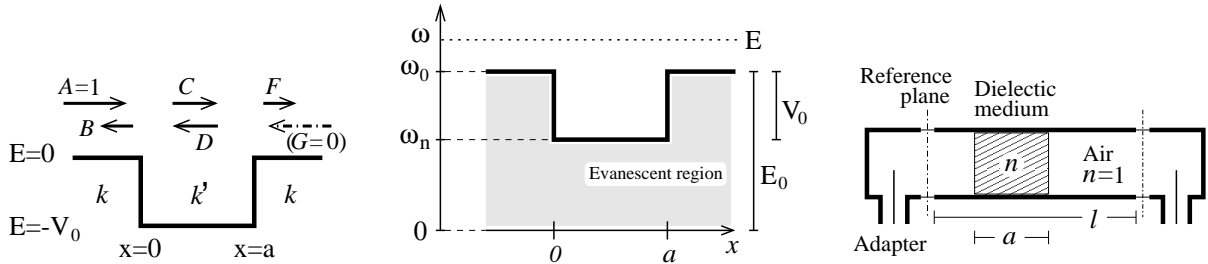}
\caption{Energy levels of the quantum well, coefficients and wave numbers of the wave function (left);
microwave analogy using wave guide sections with different cut--off 
frequencies~$\omega_0$, $\omega_n$ (center); and the experimental set--up (right): a wave guide 
of fixed length $l$ is connected to a network analyzer by coaxial adaptors. It partially 
contains a dielectric medium of refractive index $n$, which represents the quantum well.\label{topf}}
\end{figure}

In order to analyze a particle scattered by a potential well, the Schr\"odinger equation 
must be solved for a potential as sketched in Fig.~\ref{topf} (left). However, the analogy 
between the Schr\"odinger and the Helmholtz equation (\ref{schroedinger}) allows to 
examine the same process in an experiment with electromagnetic waves, too. 
\begin{equation}  \label{schroedinger}
   \left[\frac{d^2}{dx^2} + \frac{2m}{\hbar^2}\left(E-V(x)\right)\right] \psi(x) = 
0 \:,\quad
   \left[\frac{d^2}{dx^2} + \frac{1}{v^2}\left(\omega^2\!-\omega_c^2(x)\right)\right] 
\phi(x) = 0
\end{equation}
In contrast to quantum mechanics, the phase and the absolute value of the transmitted 
electromagnetic wave are measurable. Identical boundary conditions for the 
electromagnetic field~$\phi$ (representing the E or H field) and the wave 
function~$\psi$ lead to an analogous solution of the propagation problem~\cite{hanna}.

The energy level $V(x)$ of the quantum well can be constructed in a microwave 
experiment by wave guide sections with different cut--off frequencies $\omega_c(x)$, 
see Fig.~\ref{topf} (center). Applying the analogy, 
the energy baseline of the well is shifted by a constant value $E_0=\hbar \omega_0$, 
which corresponds with the cut--off frequency $\omega_0$ of the first wave guide 
section. Using the following Ansatz for the wave function, see Fig.~\ref{topf} (left):
\begin{equation} \label{ansatz_wellenfunktion}
\begin{array}{lll}
           & A \, e^{ikx} + B \, e^{-ikx}\:, & x \le 0\\
 \psi(x) = & C \, e^{ik'\!x} + D \, e^{-ik'\!x}\:, & 0 < x < a\\
           & F \, e^{ik(x\!-\!a)} + G \, e^{-ik(x\!-\!a)}\:, & a \le x  
\end{array}
\end{equation}
This leads for energies $E>E_0$ to a wave propagation with the real wave numbers
\begin{equation} \label{wellenzahlen}
  k = \frac{1}{\hbar}\sqrt{2m(E-E_0)}  \quad {\rm and} \quad
  k' = \frac{1}{\hbar}\sqrt{2m(E + V_0 - E_0)}  \;.
\end{equation}
Boundary conditions for the wave functions and their first derivatives at $x=0$ and 
$x=a$ determine the unknown coefficients $A$, $B$, $\ldots$ $G$ of (\ref{ansatz_wellenfunktion}), see~\cite{merzbacher}.
Our definition of $\psi$ for $x \ge a$ implies, that the complete phase shift inside the well 
occurs only in the coefficients $F$ and $G$. 
Assuming an incident wave at $x=0$, we set $A=1$ and $G=0$ and find
\begin{equation}
  F = \left[\cos k'a - \frac{i}{2} \!\left(\!\frac{k'}{k}\!+\!\frac{k}{k'}\!\right)\!\sin k'a\right]^{-1} .
\end{equation}
Then, the complete phase 
shift of the transmitted wave at $x=a$ becomes
\begin{equation} \label{phase}
  \varphi = \arg(F)=\arctan \!\left( \frac{1}{2}\!\left(\!\frac{k'}{k}\!+
\!\frac{k}{k'}\!\right) \tan k'a \right) .
\end{equation}
This formula is valid for both quantum mechanical and also electromagnetic wells, 
while the phase time $\tau_\varphi = \frac{d\varphi}{d\omega} = \frac{d\varphi}{d k} \frac{d k}{d \omega}$
depends on the dispersion relation $k(w)$ of the well in question.

\section*{Electromagnetic scattering in wave guides}

Inside the wave guide the wave numbers obey the dispersion relations
\begin{equation}  \label{wellenzahlen2}
  k (\omega) = \frac{1}{c}\sqrt{\omega^2\!-\!\omega_0^2} \:, \quad 
  k'(\omega) = \frac{n}{c}\sqrt{\omega^2\!-\!\omega_n^2} \:,
%  k'_0   =  k'(\omega_0)  % = \frac{n}{c}\sqrt{\omega_0^2\!-\!\omega_n^2}
\end{equation}
where $c/n=v$ is the phase velocity in the medium. 
In the experimental set--up a rectangular wave guide of length $l=250$~mm is used, 
which is partially filled by a dielectric medium of refractive index $n$, 
Fig.~\ref{topf} (right). 
The cut--off frequencies of the empty and the 
filled wave guide sections are $\omega_0 = \pi c/b$ and
$\omega_n =\pi c/n b$ respectively, where $b$ is the width of the wave guide.
The used X--band wave guide ($b=22.86$~mm) filled with Teflon ($n=\sqrt{2.05}$) has 
the cut--off frequencies $f_0 = \omega_0/2\pi = 6.56$~GHz and $f_n = 4.58$~GHz.
Thus, the energy levels of the quantum well correspond to $E_0=\hbar\omega_0 = 27.1~\mu$eV, 
$V_0 = \hbar(\omega_0\!-\!\omega_n) = 8.2~\mu$eV.
The phase time $\tau_\varphi$, which describes the propagation of the maximum 
of a wave packet in the stationary phase approximation, is 
\begin{equation} \label{phasetime}
  \tau_\varphi = \frac{d\varphi}{d\omega} = \frac{a\,\omega}{c^2 k} \frac{2 n^2 k^2(k'^2 \!\!+\! k^2) - (k'^2 \!\!-\! k^2)\, k'^2_0 \sin(2k'a)/k'a}{4k^2 k'^2 + \left(k'^2 \!\!-\! k^2\right)^2\sin^2 \! k'a} 
\end{equation}
with constant $k'^2_0=k'^2 \!\!-\! n^2 k^2$. The high frequency limit\footnote{The high frequency limit 
of the refractive index $n(\omega)$ is 1, independent of the dielectric medium. So, the term $(k'^2\!-\!k^2)^2$ in 
the denominator of (\ref{phasetime}) becomes constant and can be neglected with respect to $4k^2k'^2$.}
 of (\ref{phasetime}) for $k$, $k'\to\infty$ is $\tau_\varphi = a \frac{\omega}{k} \frac{n^2}{c^2}$. 
With the phase velocity $v_{\rm ph}=w/k$ and the group velocity $v_{\rm gr} = a/\tau_\varphi$, 
Eq.~(\ref{phasetime}) results in the well known relationship $v_{\rm gr} v_{\rm ph} = (c/n)^2$ 
for wave guides filled with a dielectric medium.

\begin{figure}[hbt]
 \begin{center}
  \includegraphics[width=0.41\textwidth]{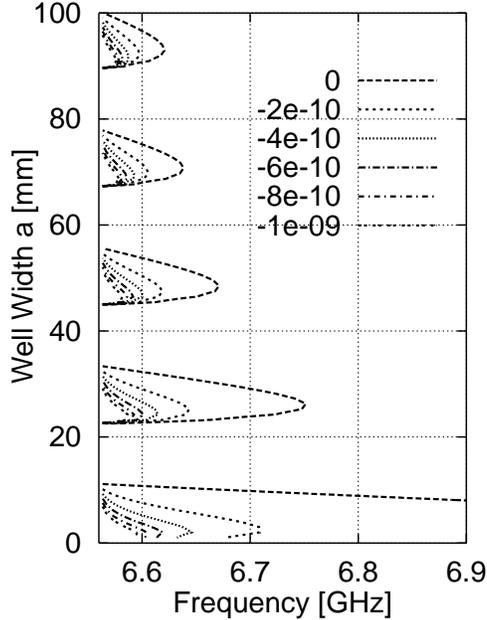}
 \end{center}
 \vspace*{-0.5cm}
 \caption{Regions of negative phase time $\tau_\varphi$ in seconds in a Teflon well according to Eq.~(\ref{phasetime}): 
depending on frequency~$f$ and well width~$a$, negative phase times between 0 and -1~ns occur 
inside the marked regions. At the left side the diagram starts with the cut--off frequency 
of the unfilled wave guide $6.56$~GHz. Below this frequency value only evanescent modes can 
propagate through the wave guide. Only certain widths $a$ yield negative phase times. 
The longer the well, the smaller the frequency band in which negative phase times appear. 
The region of negative phase time in the vicinity of $a=0$ outreaches until approximately $7.6~$GHz.\label{bedingung}}
\end{figure}

In Fig.~\ref{bedingung} we plotted regions of negative phase time depending on the well width $a$ 
and the frequency $f$ using Eq.~(\ref{phasetime}). The frequency range was chosen 
from the cut--off frequency of the empty wave guide $6.56$ up to $6.9~$GHz.

\section*{Measurement at Teflon wells}

To verify the predictions we filled tight--fitting pieces of Teflon into the wave guide. 
Some widths of the pieces fulfilled the condition for negative phase time laying inside the marked regions,
 while other pieces should not show negative phase times because their widths lay in between.

\begin{figure}[hbt]
 \begin{center}
  \includegraphics[width=0.4\textwidth]{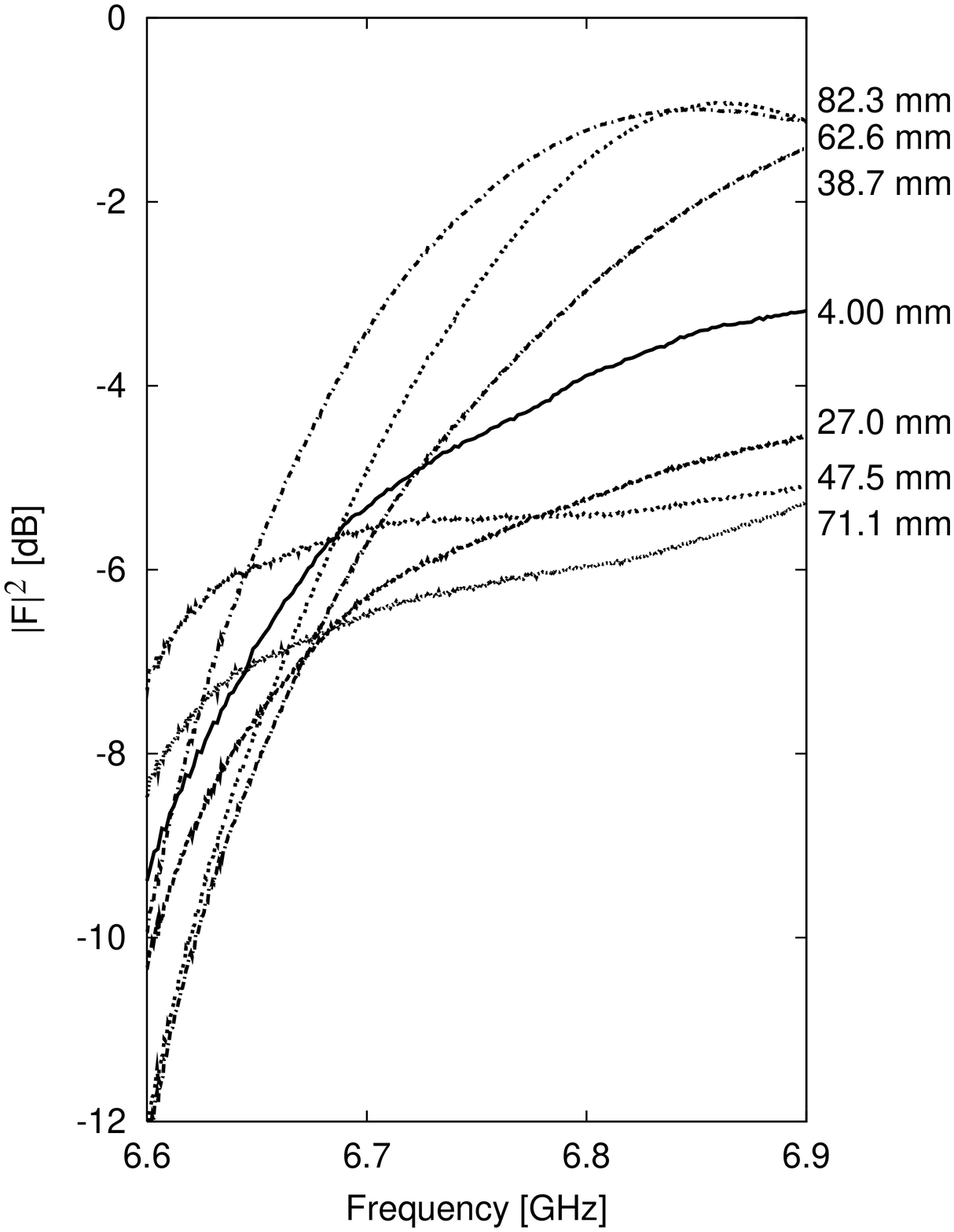}
  \includegraphics[width=0.4\textwidth]{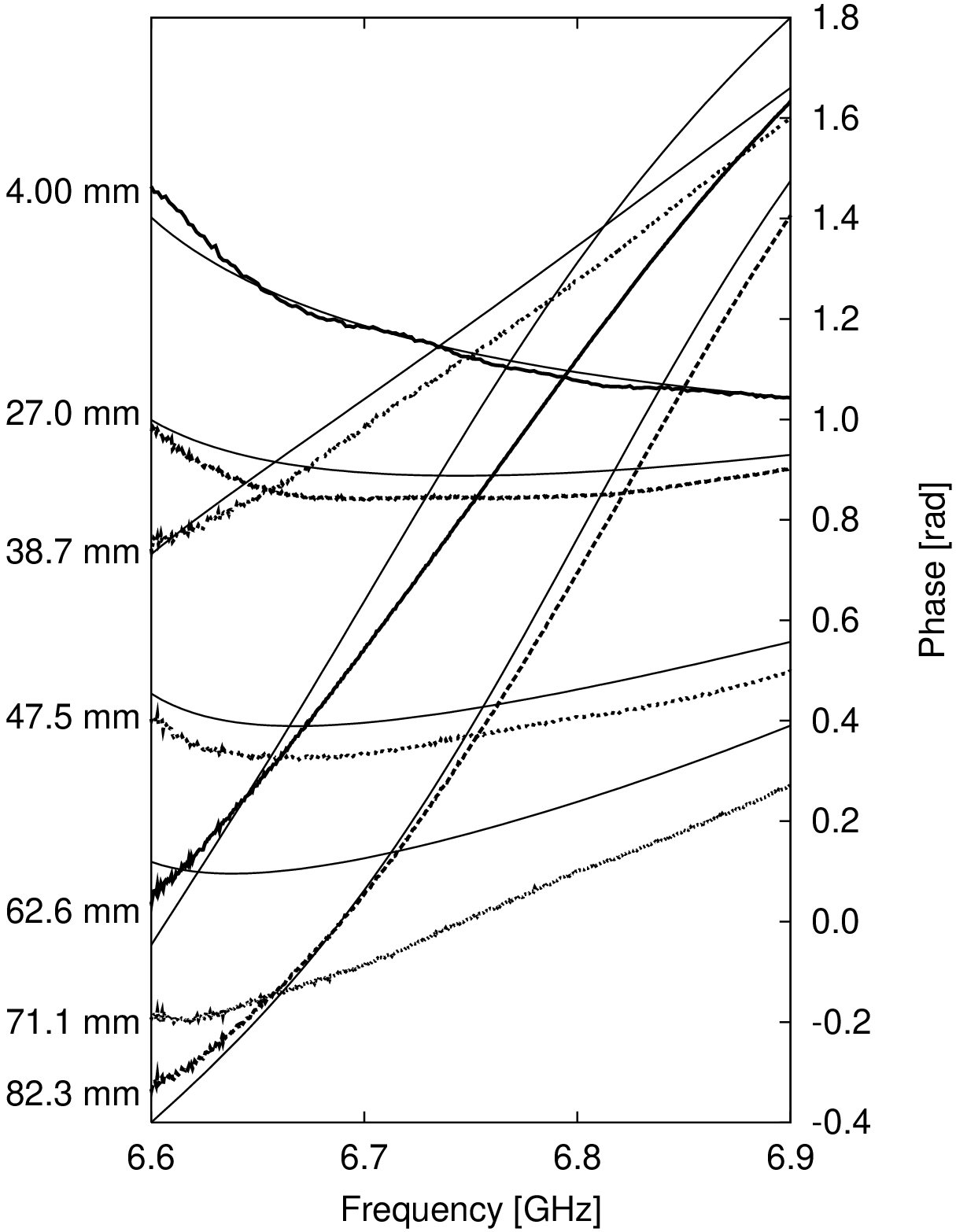}
 \end{center}
 \vspace*{-0.5cm}
 \caption{Absolute value $|F|^2$ and phase $\varphi$ of the transmission coefficient $F$ vs frequency for different 
widths of Teflon pieces. The absolute values are attenuated due to reflections at the discontinuities $x=0$,~$a$ 
and by cut--off effects. 
For some widths ($a=4.0, 27.0, 47.5$ and $71.1$~mm) the phase $\varphi$ falls with increasing frequency. Thin 
lines represent the theoretical phase progression~(\ref{phase}). To achieve a clear representation, the phase 
curves are arranged in ascending order of the well widths.\label{phasenmessung}}
\end{figure}

For each piece of Teflon, the scattering parameter $S_{21}$ for the transmission of the partially filled wave guide, 
which is defined as ratio of the transmitted to the incident wave, was determined in the frequency domain. 
The measurements were performed by a network--analyzer HP-8510, which allows an asymptotic measurement 
by eliminating undesired influences of the electrical connections by a calibration to two 
reference planes, Fig.~\ref{topf} (right). 
The measured transmission $S_{21}$ has to be corrected by a factor describing 
the change of phase inside the unfilled wave guide 
sections of total length $l-a$:
\begin{equation}  \label{korrektur}
  F = S_{21} \cdot e^{-ik(l-a)}
\end{equation}
With this operation, the reference planes in Fig.~\ref{topf} (right) are 
shifted to the surfaces of the Teflon block.
We used the measured transmission function of the
unfilled wave guide $S^{\rm ref}_{21}$ as a reference. Thus, the total phase shift 
inside the medium, which should comply with (\ref{phase}), is obtained from the 
measured data by 
\begin{equation}
  \varphi = {\rm arg}\left(S_{21}\right)-(l\!-\!a)\cdot{\rm arg}\!\left(S^{\rm ref}_{21}\right)  .
\end{equation}
Figure~\ref{phasenmessung} shows $|F|^2$ and the phase $\varphi$ of $F$ for 
microwave transmission as a function of frequency across the well for different widths $a$. For $a=38.7, 62.6$ 
and $82.3$~mm the phase increases with increasing frequency, while for 
$a=4.0, 27.0, 47.5$ and $71.1$~mm the phase decreases near the cut--off frequency.

\begin{figure}[hbt]
 \begin{center}
  \includegraphics[width=0.40\textwidth]{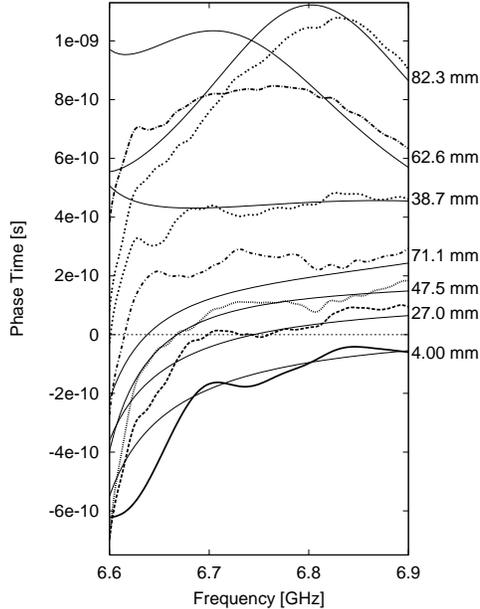} 
 \vspace*{-0.5cm}
 \end{center}
 \caption{Phase time $\tau_\varphi$ obtained from numerical derivation of the measured phase data
of Fig.~\ref{phasenmessung} and theoretical phase times (thin lines) calculated 
by (\ref{phasetime}). The curves for the well widths $a=4.0, 27.0, 47.5$, and $71.1$~mm show 
negative phase times at frequencies near the cut--off frequency of the 
unfilled wave guide. The non--vanishing resistance of real wave guide walls, especially 
near the cut--off frequency, may be in charge of the experimental deviations from the theoretical results.\label{phasenzeiten}}
\end{figure}

The phase time $\tau_\varphi = (2\pi)^{-1} d\varphi/df$ was calculated from 
the measured data by numerical derivation. Figure~\ref{phasenzeiten} presents 
the results for the phase time for transmission of the different Teflon pieces. 
For the wells widths $a=4.0, 27.0, 47.5$ and $71.1$~mm negative phase times appear, while
the other wells show the normal behavior of a positive phase progression.
The frequency intervals with negative phase times are in a fair agreement with 
the predicted intervals in Fig.~\ref{bedingung}.

\section*{Measurement at Perspex wells}

\begin{figure}[hbt]
 \begin{center}
  \includegraphics[width=0.41\textwidth]{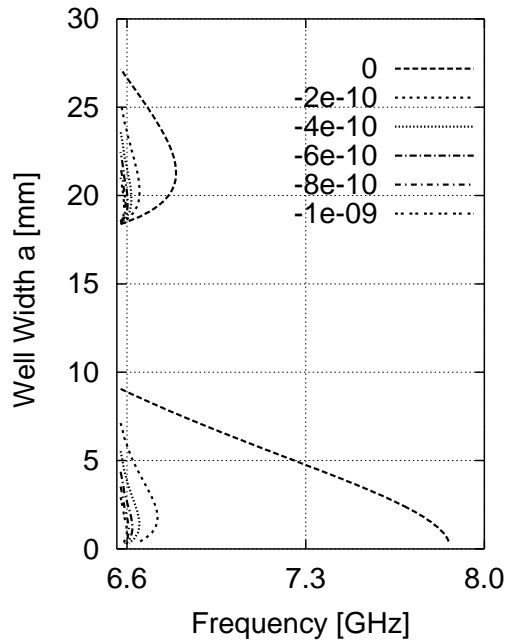}
 \end{center}
 \vspace*{-0.5cm}
 \caption{Regions of negative phase time $\tau_\varphi$ in seconds for wells made of Perspex according to Eq.~(\ref{phasetime}). The measurements were performed for the frequency interval 
 between 6.6 and 8.0~GHz and for the well widths $a=6$, 18, and 24~mm. The wells $a=6$ and $24$~mm fulfil the condition for negative phase time.\label{bedingung2}}
\end{figure}

\begin{figure}[hbt]
 \begin{center}
  \includegraphics[width=0.45\textwidth]{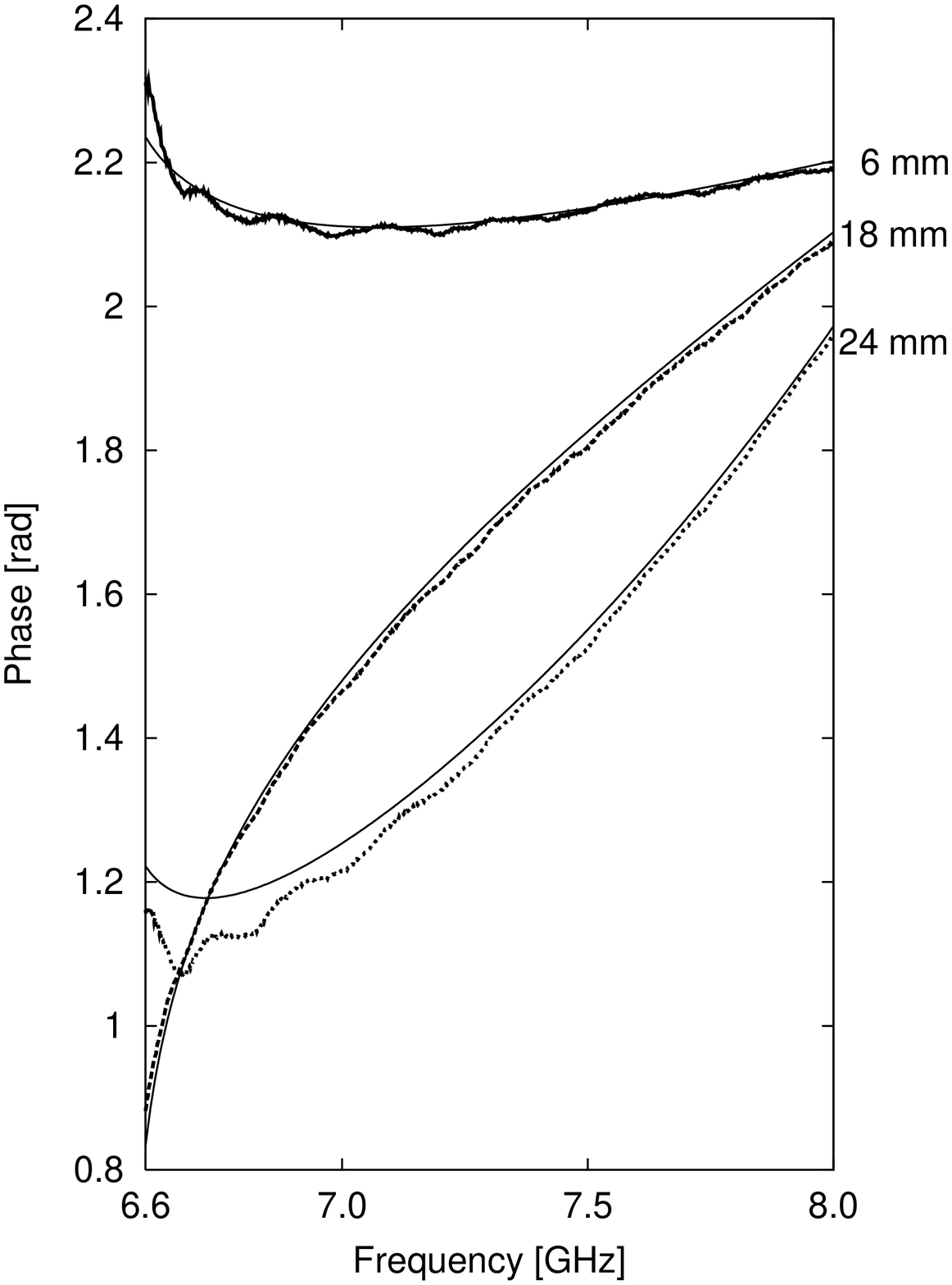}
  \includegraphics[width=0.45\textwidth]{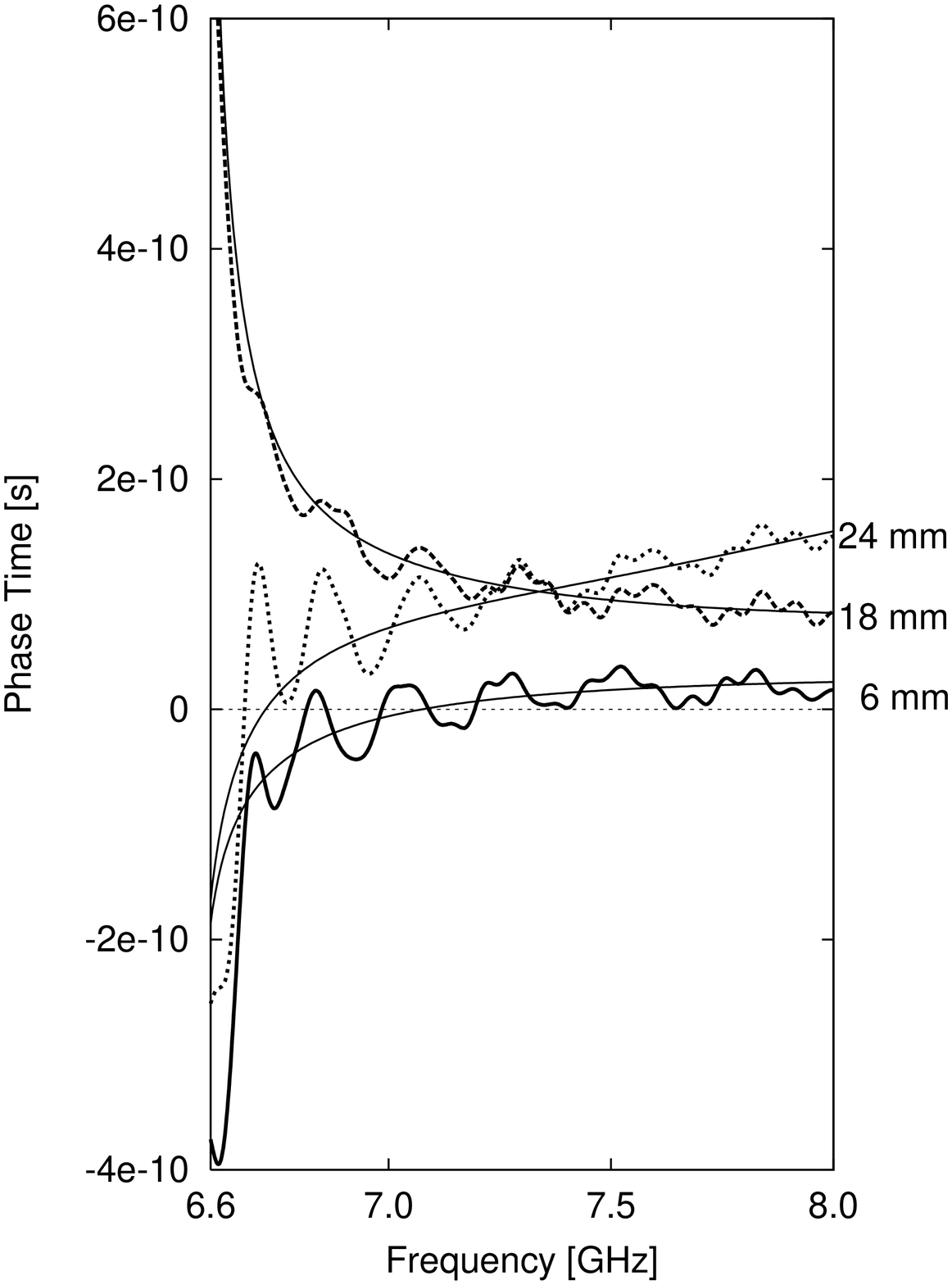}
 \end{center}
 \vspace*{-0.5cm}
 \caption{Measured phase (left) and deduced phase time (right) for the Perspex wells. The phase progression is negative 
up to 7.1~GHz for the well $a=6$~mm and up to 6.7~GHz for $a=24$~mm. The oscillations around the theoretical curves 
(thin lines) may be caused by an impedance mismatch between the wave guide and the connectors. The well width $a=18$~mm 
lays outside the marked regions of Fig.~\ref{bedingung2}, here the phase shows a normal positive progression.\label{plexibilder}}
\end{figure}

To add further credibility to the measured negative phase times we
modified the well depth $V_0$ by using Perspex as an alternative dielectric medium with  
$n=1.6$, $f_n=4.10$~GHz and $V_0=10.2~\mu$eV. According to Eq.~(\ref{phasetime}), negative phase times
are expected now for smaller well widths but for broader frequency bands, Fig.~\ref{bedingung2}. 
Measurements are performed from 6.6 to 8.0~GHz and for well widths $a=6$, 18, and 24~mm, 
the wells $a=6$ and $24$~mm should show negative phase times.
The measured phase progression in Fig.~\ref{plexibilder} (left) and the deduced phase times (right) are also in a good agreement 
with the theoretical predictions for the Perspex wells (thin lines). Although the well 
width $a=18$~mm lay close to a region of negative phase time, the measured phase progression 
of the well is clearly positive. This demonstrates how accurately the appearance of negative phase time 
depends on the width of the potential well.

\subsection*{Conclusions}

The analogy between quantum mechanical and electromagnetic scattering has been investigated with microwaves.
Theoretical studies~\cite{wang} predicted negative phase times for certain well widths if the energy of an incident particle 
is less than half the well depth. Actually, our measurements at two different well depths show the predicted 
negative phase times. For the first time, we demonstrate the effect of negative phase time for potential wells 
in a microwave analogy experiment.

\subsection*{Acknowledgements}

We are grateful to C.-F.~Li and Q.~Wang for providing their theoretical study prior to publication, and P.~Mittelstaedt for 
discussions on theory of quantum mechanical scattering.

\end{document}